\documentstyle[12pt]{article}
\title{\LARGE  Fermionic Integrals and Analytic Solutions for
Two-Dimensional Ising Models} \author{\large V.N. Plechko \\[3mm]
\em  Bogoliubov Laboratory of Theoretical Physics, \\
\em  Joint Institute for Nuclear Research, JINR-Dubna, \\
\em  141980 Dubna, Moscow Region, Russia }
\begin{document} \date{}
\newcommand{\ba}{\begin{array}}
\newcommand{\ea}{\end{array}}
\newcommand{\bea}{\begin{eqnarray}}
\newcommand{\eea}{\end{eqnarray}}
\newcommand{\sta}{\stackrel}
\newcommand{\ovl}{\overline}
\newcommand{\eee}{\mbox{e}}
\newcommand{\triex}{\mbox{\hspace{3ex}}}
\newcommand{\lla}{\longleftarrow}
\newcommand{\lra}{\longrightarrow}
\newcommand{\edoc}{\end{document}}
\newcommand{\Pfaff}{\mbox{Pfaff}}
\newcommand{\ch}{\mbox{ch}\,}
\newcommand{\sh}{\mbox{sh}\,}
\newcommand{\th}{\mbox{th}\,}
\newcommand{\pal}{\partial}
\newcommand{\spsigma}{\ba[t]{c} \mbox{Sp} \vspace{-1ex} \cr
\mbox{$\scriptstyle{(\sigma)}$} \ea }
\newcommand{\spsigmamn}{\ba[t]{c} \mbox{Sp} \vspace{-1ex} \cr
\mbox{$\scriptstyle{(\sigma_{mn})}$} \ea }
\newcommand{\spursigmamn}{\ba[t]{c} \mbox{Sp} \vspace{-1ex} \cr
\mbox{$\scriptstyle{(\sigma_{mn})}$} \ea }
\newcommand{\spura}{\ba[t]{c} \mbox{Sp} \vspace{-1ex} \cr
\mbox{$\scriptstyle{(a)}$} \ea }
\newcommand{\spurasigma}{\ba[t]{c} \mbox{Sp} \vspace{-1ex} \cr
\mbox{$\scriptstyle{(\sigma\,|\,a)}$} \ea }
\newcommand{\spuramn}{\ba[t]{c} \mbox{Sp} \vspace{-1ex} \cr
\mbox{$\scriptstyle{(a_{mn})}$} \ea }
\newcommand{\spurbmn}{\ba[t]{c} \mbox{Sp} \vspace{-1ex} \cr
\mbox{$\scriptstyle{(b_{mn})}$} \ea }

%
\maketitle
\begin{abstract}  We review some aspects of the fermionic interpretation
of the two-dimensional Ising model. The use is made of the notion of the
integral over the anticommuting Grassmann variables. For simple and more
complicated 2D Ising lattices, the partition function can be expressed as
a fermionic Gaussian integral. Equivalently, the 2D Ising model  can be
reformulated as a free-fermion theory on a lattice. For regular lattices,
the analytic solution then readily follows by passing to the momentum space
for fermions. We also comment on the effective field-theoretical
(continuum-limit) fermionic formulations for the 2D Ising models near the
critical point. Key words: Ising model, anticommuting Grassmann variables,
critical point. PACS number(s): 64.60.Cn; 75.10.-b; 05.30.Fk.\footnote{ \
Published in: J. Phys. Studies (Ukr), Vol.\ 1, No.\ 4 (1997) p.\ 554-563.}
\end{abstract}

%
%
%

\section{ Introduction.}

The two-dimensional (2D) Ising model in a zero magnetic field was first
solved by Onsager in his celebrated paper [1]. This remarkable solution
have already contributed much to our understanding of the nature of the
second-order phase transitions in magnets [2-6]. The original algebraic
method used by Onsager, however, was enough complicated. The modern
approaches to the 2D Ising model are based merely on the fermionic
interpretation of this model [7-23].  The fermionic structures in the 2D
Ising model were first recognized within the transfer-matrix and
combinatorial considerations [7,8]. It was realized later on that the
notion of the integral over anticommuting Grassmann variables due to
Berezin [9,10] is a powerful tool to analyze the $2D$ Ising models [10-23].
Berezin himself was the first to apply the anticommuting variables to
evaluate the partition function of the 2D Ising model on rectangular
lattice [10]. His method was based on combinatorial considerations [10].
The fermionic approaches to the 2D Ising model have been developed later on
by many authors [11-23]. In the forthcoming discussion we will merely
follow a simple fermionic interpretation of the $2D$ Ising model introduced
in [17-19]. The approach is based on integration over the anticommuting
Grassmann variables and the mirror-ordered factorization principle for the
2DIM density matrix.  The method do not involve the traditional
transfer-matrix or combinatorial considerations.  Schematically, we
have:  \bea Q\,=\,\spsigma\,Q\,(\sigma) \to \spurasigma\,Q\,
(\sigma\,|\,a)\to \spura\,Q\,(a)\,=\,Q\,. \label{eq1}\eea   Here we start
with the original spin partition function, $Q$, and introduce, in a special
way, a set of new anticommuting Grassmann variables $(a)$ thus passing to a
mixed $(\sigma\,|\,a)$ representation for $Q$.  Eliminating the Ising spin
variables in this mixed $(\sigma\,|\,a)$ representation, we obtain a purely
fermionic expression for the same partition function $Q$. The partition
function appears, in final form, as a Gaussian fermionic integral.
Equivalently, the 2D Ising model is represented as a free-fermion field
theory on a lattice [17-19]. For the homogeneous (translationally
invariant) lattices the analytic solution for the partition function and
free energy then readily follows by passing to the momentum space for
fermions by means of Fourier substitution. In particular, this gives a
few line derivation of the Onsager result [17]. The exact solution for a
finite lattice on a torus (periodic boundary conditions in both directions)
also follows within suitable modification of the fermionization procedure
[18]. The decomposition of the partition function on a torus into a
characteristic sum of four fermionic integrals follows here from a general
rule of transposition of two Grassmann variable functions derived in
Appendix A of [18]. The Ising models settled on the lattices with
complicated local structures have being analyzed by factorization method
within the spin-polynomial interpretation of the problem [19]. We will pay
more attention to this last case in sections 3-7.  For application of the
results of fermionic analysis [19] in studies of regularly diluted Ising
ferromagnets also see [25,26]. It is also important that the method works
in fact for the most general inhomogeneous distribution of the coupling
parameters over the lattice bonds [17].  This may be of interest in studies
of the disordered Ising models [20-23]. The Gaussian fermionic
representations has been recently constructed as well for the
inhomogeneous two-dimensional dimer problems [24].

In the straightforward variant of the method [17] we introduce Grassmann
variables already at the first stages in order to decouple the local bond
Boltzmann weights into separable factors called Grassmann factors
(GFs). We then combine GFs with the same spin variables into separable
groups all over the lattice and then sum over spin states in each group
independently, thus passing to a purely fermionic representation for $Q$.
In general, the factorization idea we apply resembles the idea of insertion
of the Dirac unity, $\Sigma\,|\,a\left> \right<a\,|=1$, in quantum
mechanics. However, the situation in our case is more complicated since we
deal here with Grassmann factors which are neither commuting nor
anticommuting, in general, with each other.  The key point of all the
construction is then the mirror-ordering procedure for the global products
of GFs [17]. This ordering procedure enables us to keep nearby few relevant
factors with the same spin variable in the process of fermionization, which
is the necessary condition in order we can actually eliminate spin degrees
of freedom. The ordering procedure for global product of GFs is the most
important part of the solution [17-19]. It might be worthwhile also
mentioning that the lacking of a suitable ordering procedure for GFs in
the 3D case is a serious obstacle to deal with the 3D Ising model within
fermionic factorization method. The same holds true for the 2D Ising model
in a non-zero magnetic field.

If we start with factorization of the local bond Boltzmann weights, as
we do for the standard rectangular lattice in [17,18], we obtain the
fermionic representation for $Q$ with four variables per site. Starting by
factorization of the bond weights, the number of fermionic variables
per site will be even larger for more complicated lattices. An important
modification of the method was introduced in [19], where we start with
factorization of the local cell weights presented by three-spin
polynomials. The spin-polynomial interpretation for the 2D Ising models
arises if we multiply few local weights forming elementary cell in $Q$.
Effectively, this reduces the number of fluctuating degrees of freedom
already on the level  of the spin-variable formulation of the problem,
prior to fermionization.  Respectively, the Gaussian integral for $Q$
appears here with only two fermionic variables per site, see (\ref{qccc})
below, which provides essential simplifications in the forthcoming
analysis [19]. For a set of 2D Ising models, including rectangular,
triangular and hexagonal Ising lattices as the simplest cases, this gives
analytic solution in terms of the parameters of the three-spin
polynomial characterizing elementary cell. The exact lattice integral
(\ref{qccc}) is also a suitable starting point to pass to the effective
continuum-limit formulation near $T_c$.  The continuum-limit formulation
corresponds to the low-momentum sector of the exact lattice theory
responsible for the critical-point singularities and the large-distance
behaviour of correlations near $T_c$.  The resulting theory is the massive
2D Majorana theory, with mass vanishing at $T_c$. By doubling of fermions
in Majorana representation, we can pass as well to the 2D Dirac field
theory of charged fermions. Below we comment on some aspects of
the fermionic interpretation of the 2D Ising model in more detail.  The
fragments of present discussion have been reported also in [27,28]. In the
next section we outline shortly the basic rules of fermionic integration.

%
%
\section{ Grassmann variables.}

We remember that Grassmann variables are the purely anticommuting fermionic
symbols.  Given a set of Grassmann variables $a_1,a_2, ..., a_N$, we have
$a_ia_j+ a_ja_i =0$, $a_{j}^{2}=0$. The Beresin's rules of integration for
one variable are [9]:
\bea
\int da_j \cdot a_j = 1\,, \triex\triex
\int da_j \cdot 1   = 0\,.
\label{gr1}
\eea
In the multidimensional integral, the differentials $da_1, da_2,...\,,
da_N$ are again anticommuting with each other and with the variables.  The
basic relations of the grassmannian analysis concern the Gaussian fermionic
integrals [9,10]. The Gaussian integral of the first kind is related to the
determinant:
\bea
\int\,\prod\limits_{j=1}^{N}\,da_{j}^{*}da_{j}\,\exp\left(
\sum\limits_{i=1}^{N}\sum\limits_{j=1}^{N}a_{i}A_{ij}a_{j}^{*}
\right)\,=\,
\det\hat{A}\;,
\label{gr2}
\eea
where $\{a_{j},a_{j}^{*}\}$ is a set of completely anticommuting Grassmann
variables, the matrix in the exponential is arbitrary. The fermionic
exponential here is assumed in the sense of its series expansion. The
series terminates at some stage due to the property $a_{j}^{\,2}=0$. The
correspondent finite polynomial for the exponential in (\ref{gr2}) can be
obtained also by multiplying the elementary factors $\exp\,(a_iA_{ij}a_{j}
^{\,*}) = 1 + a_{i}^{}A_{ij}^{} a_{j}^{\,*}$. The appearance of determinant
in (3) is due to the known interrelations between fermionic algebra and
determinant combinatorics. By convention, the variables $a_{j}^{}$ and
$a_{j}^{\,*}$ can be considered as complex conjugated fermionic  fields,
otherwise these are simply independent variables. The Gaussian integral of
the second kind, for real fermionic fields, is related to the Pfaffian of
the associated skew-symmetric matrix:
\bea
\int\,da_N\,...\,da_2da_1\,\exp\left(
\frac{1}{2}\,\sum\limits_{i=1}^{N}\sum\limits_{j=1}^{N}
a_{i}A_{ij}a_{j}\right)\,=\,\mbox{Pfaff}\,\hat{A}\,,
\hspace{3ex} A_{ij}=-A_{ji}\,.
\label{gr3}
\eea
The Pfaffian is some combinatorial polynomial in elements $A_{ij}^{}$
known in mathematics for a long time. Under another name, the Pfaffian
is also well known in physics since the Pfaffian combinatorics is
identical with that of the fermionic version of Wick's theorem. For any
skew-symmetric matrix $(A_{ij} = - A_{ji})$ we have:
\bea
\det \hat{A} =(\,\mbox{Pfaff}\, \hat{A}\,)^{\,2}\,.
\label{detpf}
\eea
This identity can be most  easily proved just in terms of the fermionic
integrals like (\ref{gr2}) and (\ref{gr3}). The fermionic averages (Green's
functions) correspondent to the integrals (\ref{gr2}) and (\ref{gr3}) can
be defined in a natural way. The linear superpositions of Grassmann
variables are again Grassmann variables and it is possible to make a linear
change  of variables in the fermionic integrals. As compared with the rules
of commuting analysis, the only difference is that Jacobian will now appear
in the inverse power [9,10]. [It might be also interesting to compare
(\ref{gr2}) with the analogous Gaussian integral over the commuting
complex-variable fields in common analysis].  In physical applications, the
new fermionic variables of integration are often introduced by Fourier
substitution (transformation to the momentum space).

%
%
\section{  The spin-polynomial interpretation.}

In order to construct the fermionic representation for $Q$ with the minimal
number of fermionic variables per site, we start with the spin-polynomial
formulation of the problem [19]. This can be most simply illustrated by an
example of the Ising model on a triangular lattice. The triangular lattice
can be viewed as a rectangular one with an additional diagonal interaction
introduced in each rectangular cell. Let lattice sites on the
corresponding rectangular net be marked by integer Cartesian coordinates
$\,mn\,$, with $m=1,2,...,L$, $n=1,2,..., L$ running in horizontal and
vertical directions, respectively. Here $L$ is the lattice length, $N=L^2$
is the total number of sites and spins in a lattice. At final stages we
assume $L^2\to \infty$. With each lattice site we associate the Ising spin
variable $\sigma_{mn}=\pm1$. The hamiltonian of the Ising model on a
triangular lattice is then given as follows:
\bea
-\beta\,H\,(\sigma)=\sum\limits_{mn}^{}\;[\;b_1\,\sigma_{mn}
\sigma_{m+1n}+ b_2\,\sigma_{m+1n}\sigma_{m+1n+1}+ b_3\,\sigma_{mn}
\sigma_{m+1n+1}\,]\,, \;\;\; \beta=1/kT\,,
\label{ham1}
\eea
where $b_{\alpha} = J_{\alpha}/ kT$ are the dimensionless coupling
constants, $J_{\alpha}$ are the magnetic exchange energies, and $kT$ is the
temperature in energy units. To be definite, in what follows we will merely
keep in mind the ferromagnetic case, $b_{\alpha}>0$, though this
restriction is not essential until we pass to the continuum limit
formulations. The partition function and the free energy per site are:
\bea
Z=\sum_{(\sigma)}^{}\,\eee^{\,-\beta H(\sigma)}\,,\triex
-\beta\,f_{\,Z}=\, \lim\limits_{N\to\infty}^{}\frac{1}{N}\ln\,Z\,,
\label{zf}
\eea
where the sum is taken over $2^N$ spin configurations provided by
$\sigma_{mn} =\pm1$ at each site.

Noting the identity for the typical bond weight: $\eee^{\,b\, \sigma
\sigma'}=\cosh\,b\,+\,\sinh\, b\cdot\sigma \sigma'$, which readily
follows from $(\sigma\sigma')^{2}=+1$, we find:
\bea
Z=\,(2\,\cosh\,b_1\,\cosh\,b_2\,\cosh\,b_3)^{\,N}\,Q\,, \triex
N=L^{2}\to\infty\,,
\label{zq}
\eea
where $Q$ is the reduced partition function:
\bea
Q = \spsigma\{\,\prod\limits_{mn}\,(\,1+t_1\,\sigma_{mn}\sigma_{m+1n})\,
(1+t_2\,\sigma_{m+1n}\sigma_{m+1n+1})\,
(1+t_3\,\sigma_{mn}\sigma_{m+1n+1})\,\}\,,
\label{qspin}
\eea
where $t_{\alpha}=\tanh\,b_{\alpha}$, $b_\alpha=J_\alpha/kT$, and
$Sp_{(\sigma)}$ is the properly normalized spin averaging [19]:
\bea
\spsigma\,(...)=\prod\limits_{mn}^{}\spsigmamn\,(...) =
2^{\,-N}\sum\limits_{(\sigma)}^{}(...)\,, \;\;\;
\spsigmamn(...) = \frac{1}{2} \sum\limits_{\sigma_{mn}=\pm{1}}(...)\,,
\label{aver}
\eea
the local averaging is normalized here in such a way that $\mbox{Sp}\,(1)
=1$, also note that $\mbox{Sp}\,(\sigma_{mn})=0$.  The reduced partition
function $Q$ will be the main subject of our interest.

In order to pass to the spin-polynomial interpretation, we have to multiply
the three bond weights forming a triangular cell in $Q$.  Let us introduce
the local enumeration of sites at our rectangular net:
\bea
(\,\sigma_{1}\,|\,\sigma_{2}\,|\,\sigma_{3}\,|\,\sigma_{4}\,)
\leftrightarrow
(\,\sigma_{mn}\,|\,\sigma_{m+1n}\,|\,\sigma_{m+1n+1}\,|\,
\sigma_{mn+1}\,)\,.
\label{local}
\eea
The elementary cell in (\ref{ham1}) and (\ref{qspin}) is formed by a
triangle of spins $(\sigma_{1},\sigma_{2},\sigma_{3})_{\,mn}$. Noting that
$\sigma_{j}=\pm1$, and hence $\sigma_{j}^{\,2}=1$, we can multiply
the weights in (\ref{qspin}) making use of the properties like $\sigma_1
\sigma_2 \cdot \sigma_2\sigma_3 = \sigma_1 \sigma_3$, etc. For the product
of the three bond weights from (\ref{qspin}) we then obtain the three-spin
polynomial as is given in (\ref{p123}) below.  This polynomial can be
considered as the effective Boltzmann weight of a triangular elementary
cell taken as a whole:
\bea
\label{p123}
&& P_{123}(\sigma) = (1+t_1\,\sigma_{1}\sigma_{2})\,
(1+t_2\,\sigma_{2}\sigma_{3})\,(1+t_3\,\sigma_{1}\sigma_{3})
\nonumber \\
&&
=\,(1+t_1t_2t_3) + (t_1+t_2t_3)\,\sigma_{1}\sigma_{2} +
(t_2+t_1t_3)\,\sigma_{2}\sigma_{3} +
(t_3+t_1t_2)\,\sigma_{1}\sigma_{3}\,
\\ \nonumber
&& = \alpha_0 + \alpha_1\,\sigma_{1}\sigma_{2} +
\alpha_2\,\sigma_{2}\sigma_{3} + \alpha_3\,\sigma_{1}\sigma_{3}\,.
\eea
In the last line we write the three-spin polynomial in general
notation, for the given case of triangular lattice we have:
\bea
\alpha_0=1+t_1t_2t_3\,,\, \triex\alpha_1= t_1+ t_2t_3\,,\, \triex
\alpha_2=t_2+ t_1t_3\,,\triex \alpha_3=t_3 + t_1t_2\,,
\label{alpha}
\eea
while the case of rectangular lattice follows with $t_3=0$, which
corresponds to $b_3=0$ in the hamiltonian (\ref{ham1}). For rectangular
lattice we have: $\alpha_0=1,\, \alpha_1= t_1,\,\alpha_2= t_2,\,
\alpha_3= t_1t_2\,$. The hexagonal and more complicated lattices
can be characterized by the three-spin polynomials like (\ref{p123}) as
well, with their own sets of the $\alpha$-parameters. Thus we come to the
three-spin polynomial partition function [19]:
\bea
Q=\spsigma \prod\limits_{mn}^{}\,(\alpha_0 + \alpha_1\,
\sigma_{mn}\sigma_{m+1n}+ \alpha_2\,\sigma_{m+1n}\sigma_{m+1n+1}+
\alpha_3\,\sigma_{mn}\sigma_{m+1\,n+1}\,)\,.
\label{q123}
\eea
In what follows we assume $\alpha_0, \alpha_1,\alpha_2, \alpha_3\,$
to be arbitrary parameters. Particular Ising lattices are specified by
the choice of the $\alpha$ -parameters [19,25,26].

%
%
\section{ Free-fermion representation and analytic
\newline  \ results (lattice case).}

The three-spin polynomial partition function (\ref{q123}) can be
transformed into a fermionic Gaussian integral following (\ref{eq1}). The
starting point is factorization of the local polynomial weights in
(\ref{q123}). We need only two variables per site in the process of
fermionization. The partition function then appears as the
following Gaussian fermionic integral \cite{plech3}:
\bea
Q\, =
\int \prod\limits_{mn}^{}dc_{mn}^{*}dc_{mn}^{}\exp\,
\sum\limits_{mn}^{}\,[\,(\alpha_{0}\,c_{mn}c_{mn}^{*} -
\alpha_1\,c_{mn}c_{m-1n}^{*} - \alpha_2\,c_{mn}c_{mn-1}^{*}
\cr
-\,\alpha_{3}\,c_{mn}c_{m-1n-1}^{*}) -\alpha_{1}\,c_{mn}c_{m-1n}-\,
\alpha_{2}\,c_{mn}^{*}c_{mn-1}^{*}\,]\;,
\label{qccc}
\eea
where $c_{mn}^{}, c_{mn}^{*}$ are the totally anticommuting Grassmann
variables, two per site. The free-fermion representation (\ref{qccc}) is
exact. The integral (\ref{qccc}) is equivalent to the original partition
function (\ref{q123}) up to the boundary effects, which are inessential as
$L^2\to\infty$.  The explicit evaluation of the integral (\ref{qccc}) can
be performed by passing to the momentum space by means of Fourier
substitution:
\bea
c_{mn}^{}= \frac{1}{L}\sum\limits_{pq}^{} c_{pq}^{}\,
\eee^{\,i\,\frac{2\pi}{L}(mp+nq)}\,, \triex
c_{mn}^{\,*}= \frac{1}{L}\sum\limits_{pq}^{} c_{pq}^{\,*}\,
\eee^{\,-\,i\,\frac{2\pi}{L}(mp+nq)}\,. \triex
\label{furccc}
\eea
Here $c_{pq}^{}, c_{pq}^{\,*}$ are the new fermionic variables of
integration. The integral (\ref{qccc}) now appears in the form:
\bea
Q = \int\limits_{}^{} \prod\limits_{pq}^{}dc_{pq}^{\,*}dc_{pq}^{}
\exp\,\sum\limits_{pq}^{}\left[\,c_{pq}^{}c_{pq}^{\,*}\left(
\alpha_0 - \alpha_1\,\eee^{\,i\,\frac{2\pi p}{L}}
- \alpha_2\,\eee^{\,i\,\frac{2\pi q}{L}}
- \alpha_3\,\eee^{\,i\,\frac{2\pi}{L}(p+q)}\right)\right.
\cr
\left.
-\,\alpha_1\,c_{pq}^{}c_{L-pL-q}^{}\,\eee^{\,i\,\frac{2\pi p}{L}} -
\alpha_2\,c_{L-pL-q}^{\,*}c_{pq}^{\,*}\,\eee^{\,i\,\frac{2\pi q}{L}}
\right]\;.
\label{qc2}
\eea
Note that the fermionic measure transforms in a trivial way by passing from
(\ref{qccc}) to (\ref{qc2}). This is because the Jacobian of the
substitution (\ref{furccc}) is unity, as it follows from the orthogonality
of the Fourier eigenfunctions.

It is easy to see that the integral (\ref{qc2}) decouples into a  product
of simple low-dimensional integrals. Since the variables with momenta $pq$
and $L-pL-q$ interact in (\ref{qc2}), we have to single out in the
fermionic action in (\ref{qc2}) the combined $(pq\,|\,L-pL-q)$ term. Let
$S_{\,pq}$ be the term already given explicitly in the $pq$ sum above. It
then follows that the integral (\ref{qc2}) is the product of the following
independent factors:
\bea
Q_{\,pq}^{\,2}= \int D_{\,pq}\,\exp\, \big( S_{\,pq} + S_{\,L-pL-q})\,,
\triex  D_{\,pq} = dc_{L-pL-q}^{\,*}dc_{L-pL-q}^{}
dc_{pq}^{\,*}dc_{pq}^{}\,.
\label{factor3}
\eea
The elementary integral (\ref{factor3}) can be readily evaluated by
elementary tools (for instance, simply using the definitions (\ref{gr1}))
and its value is given as a factor in the product (\ref{qotwet}) below.
By comparing the fermionic measures in (\ref{qc2}) and (\ref{factor3}), it
follows that the partition function itself, $Q$, arises if we multiply the
factors $(\ref{factor3})$ over only one half of the points in the momentum
space. That is, we have to multiply the factors (\ref{factor3}) in such a
way that if the given mode $pq$ is already included into the product, then
the conjugated mode $L-pL-q$ is not to be included, and vice versa.
Respectively, the total product of factors (\ref{factor3}) over the
complete set of momentum modes, $0 \le p,q \le L-1$, will yield the
squared partition function. Thus, for the partition function
(\ref{qccc}) we find [19]:
\bea Q^{\,2} =
\prod\limits_{p=0}^{L-1}\prod\limits_{q=0}^{L-1}\,\big[\,
(\alpha_{0}^{2}+\alpha_{1}^{2}+\alpha_{2}^{2}+\alpha_{3}^{2}) -
2\,(\alpha_{0}^{}\alpha_{1}^{}-\alpha_{2}^{}\alpha_{3}^{})\,\cos
\frac{2\pi p}{L}\,
\cr
-\,2\,(\alpha_{0}^{}\alpha_{2}^{}-\alpha_{1}^{}\alpha_{3}^{})\,\cos
\frac{2\pi q}{L} -
2\,(\alpha_{0}^{}\alpha_{3}^{}-\alpha_{1}^{}\alpha_{2}^{})\,\cos
\frac{2\pi (p+q)}{L}\,\big]\;.
\label{qotwet}
\eea
Respectively, the free energy per site is:
\bea
&& -\beta f_Q = \,\lim_{L\rightarrow\infty}^{}\,
(\frac{1}{L^2}\ln Q\,)\,
\cr
&& =\frac{1}{2}\,\int\limits_{0}^{2\pi}\int\limits_{0}^{2\pi}
\frac{dp}{2\pi} \frac{dq}{2\pi}\,\ln\,\big[\,
(\alpha_{0}^{2}+\alpha_{1}^{2}+\alpha_{2}^{2}+\alpha_{3}^{2}) -
2\,(\alpha_{0}^{}\alpha_{1}^{}-\alpha_{2}^{}\alpha_{3}^{})\,\cos p\,
\cr
&& -\,2\,(\alpha_{0}^{}\alpha_{2}^{}-\alpha_{1}^{}\alpha_{3}^{})\,
\cos q\,-2\,(\alpha_{0}^{}\alpha_{3}^{}-\alpha_{1}^{}\alpha_{2}^{})\,
\cos (p+q)\,\big]\;.
\label{fotwet}
\eea
In particular, the exact solutions for the rectangular, triangular, and
hexagonal Ising lattices follow from (\ref{fotwet}) under correspondent
specifications of the parameters $\alpha_{_j}$ [19].

Here we consider for illustration the simplest case of the standard
rectangular lattice. For this lattice, we have $\alpha_0, \alpha_1,
\alpha_2, \alpha_3 \leftrightarrow 1,t_1,t_2,t_1t_2$. From (\ref{fotwet}),
the free energy is:

\bea
-\beta f_{\,Q}\,\big|\,_{\rm\,rect}^{} =\;
\frac{1}{2}\,\int\limits_{0}^{2\pi}\int\limits_{0}^{2\pi}
\frac{dp}{2\pi} \frac{dq}{2\pi}\,\ln\,\big[\,
(1+t_{1}^{2})\,(1+t_{2}^{2})
- 2t_{1}^{}\,(1-t_{2}^{2})\,\cos p\,
\cr
-\,2t_{2}^{}\,(1-t_{1}^{2})\,\cos q\,\big]\,.
\label{frec1}
\eea
The free energy  (\ref{frec1}) is associated with the reduced partition
function, $Q$, while the true partition function is $Z = (2c_1c_2)^{\,N}Q$,
see (\ref{zq}) and (\ref{qspin}) with $t_3=0$. Also we remember the
identities like $c_{\alpha}^{2}(1+t_{\alpha}^{2})=\cosh  2b_\alpha$,
$2\,t_{\alpha}c_{\alpha}^{2} =\sinh 2b_\alpha$, etc, with
$b_{\,\alpha}=J_{\,\alpha}/kT$. From (\ref{frec1}), the true free energy
per site then appears in the form:

\bea
-\beta f_{\,Z}\,\big|\,_{\rm\,rect}^{} = \ln\,2
+\,\frac{1}{2}\,\int\limits_{0}^{2\pi}\int\limits_{0}^{2\pi}
\frac{dp}{2\pi}\frac{dq}{2\pi}\,\ln\,\big[\,\cosh 2b_1\,\cosh 2b_2
- \sinh 2b_1\,\cos p
\cr
\,-\,\sinh 2b_2\,\cos q\,\big]\,.
\label{frec2}
\eea
This is the famous Onsager solution for the free energy of the 2D Ising
model on a rectangular lattice [1]. An interesting comment on the structure
of this solution follow immediately after equation (108) in [1].

As it follows from (\ref{frec1}), for rectangular lattice the critical
point is fixed by condition:

\bea
1-t_1-t_2-t_1t_2=0\,, \triex
t_\alpha=\tanh(J_\alpha/kT)\,,
\label{crit1}
\eea
equivalently, this condition can be written in the form:

\bea
\sinh(2J_1/kT)\,\sinh(2J_2/kT)\,=1\,,
\label{crit2}
\eea
which rather corresponds to the solution in the form of (\ref{frec2}). In
both formulations the ferromagnetic case is assumed, $b_1,b_2>0$,
equivalently, $t_1,t_2>0$. The free energy is known, the specific heat
can be obtained by differentiation with respect to the temperature: $C/k =
\beta^2\,(\pal^2(-\beta f)/\pal \beta^2)$, where $C/k$ is the dimensionless
specific heat, $\beta=1/kT$, and $k$ is the Boltzmann constant. The
singularity in the specific heat appears to be logarithmic near $T_c$:
$C/k \sim |\ln|\tau||$, $\tau= (T-T_c)/T_c \to0$, as it can be deduced
>from (\ref{frec1}) and/or (\ref{frec2}). Also see the discussion in the
next section.

%
%
\section{ The symmetries and the critical point. }

The symmetries provided by the above exact solution (\ref{qotwet}) and
(\ref{fotwet}), and the closely related question on the location of the
critical point, have an interesting interpretation within the
spin-polynomial language [19]. Making use of the local enumeration of sites
in the elementary cell, see (\ref{local}), the cell weight in the density
matrix in (\ref{q123}) is given as the three-spin polynomial:
\bea
P_{\,123}\,(\sigma)= \alpha_0 +\alpha_1\,\sigma_1\sigma_2 +
\alpha_2\,\sigma_2\sigma_3 + \alpha_3\,\sigma_1\sigma_3\,.
\label{pa123}
\eea
It appears also to be useful to introduce the associated three-spin
polynomial:
\bea
F_{\,123}\,(\sigma)= \alpha_0 - \alpha_1\,\sigma_1\sigma_2 -
\alpha_2\,\sigma_2\sigma_3 - \alpha_3\,\sigma_1\sigma_3\,,
\label{fa123}
\eea
and we note the following interesting identity [19]:

\bea
(F_{\,123}(\sigma))^{\,2} = (\alpha_0 - \alpha_1\,\sigma_1\sigma_2 -
\alpha_2\,\sigma_2\sigma_3 - \alpha_3\,\sigma_1\sigma_3\,)^{\,2}
\cr
\,=\,(\alpha_{0}^{2}+ \alpha_{1}^{2}+\alpha_{2}^{2}+\alpha_{3}^{2})\,-\,
2\,(\alpha_0\alpha_1-\alpha_2\alpha_3)\,\sigma_1\sigma_2
\cr\,-\,
2\,(\alpha_0\alpha_2-\alpha_1\alpha_3)\,\sigma_2\sigma_3\,-\,
2\,(\alpha_{0}\alpha_{3}-\alpha_{1}\alpha_{2})\,\sigma_1\sigma_3\,.
\label{aiden}
\eea
It is seen that the combinations of the $\alpha$-parameters occurring in
$(F_{123})^2$ are exactly the same as in the momentum modes $Q^{\,2}(p\,|
\,q)$ in (\ref{fotwet}):

\bea
Q^{\,2}\,(p\,|\,q) =
(\alpha_{0}^{2}+\alpha_{1}^{2}+\alpha_{2}^{2}+\alpha_{3}^{2}) -
2\,(\alpha_{0}^{}\alpha_{1}^{}-\alpha_{2}^{}\alpha_{3}^{})\,\cos p
\cr
\,-2\,(\alpha_{0}^{}\alpha_{2}^{}-\alpha_{1}^{}\alpha_{3}^{})\,\cos q\,-
2\,(\alpha_{0}^{}\alpha_{3}^{}-\alpha_{1}^{}\alpha_{2}^{})\,\cos (p+q)\,,
\label{amodes}
\eea
where $0\leq p,q \leq2\pi$, the limit $L^2\to\infty$ is already assumed.

It appears that the following combinations of the $\alpha$-parameters play
important role in discussing the symmetries and the critical point [19]:
\bea
\ba{llr}
\alpha_{0}^{*}=\frac{1}{2}\,(\alpha_0+\alpha_1+\alpha_2+\alpha_3)\,,\cr
\alpha_{1}^{*}=\frac{1}{2}\,(\alpha_0+\alpha_1-\alpha_2-\alpha_3)\,,\cr
\alpha_{2}^{*}=\frac{1}{2}\,(\alpha_0-\alpha_1+\alpha_2-\alpha_3)\,,\cr
\alpha_{3}^{*}=\frac{1}{2}\,(\alpha_0-\alpha_1-\alpha_2+\alpha_3)\,,
\ea
& \;\;\;
\ba{llr}
\bar{\alpha}_{0}^{}=\frac{1}{2}\,(\alpha_0-\alpha_1-\alpha_2-\alpha_3)\,,\cr
\bar{\alpha}_{1}^{}=\frac{1}{2}\,(\alpha_0-\alpha_1+\alpha_2+\alpha_3)\,,\cr
\bar{\alpha}_{2}^{}=\frac{1}{2}\,(\alpha_0+\alpha_1-\alpha_2+\alpha_3)\,,\cr
\bar{\alpha}_{3}^{}=\frac{1}{2}\,(\alpha_0+\alpha_1+\alpha_2-\alpha_3)\,.
\end{array}
\label{aeigen}
\eea
The parameters $\alpha^*$ and $\bar{\alpha}$ are in fact the eigenvalues
of the polynomials $\frac{1}{2}P_{123}$ and $\frac{1}{2} F_{123}$,
respectively. By the "eigenvalues" we mean the four numbers which takes the
polynomial as the spin variables run over their permissible values $\pm1$.
We note also the following important identity [19]:
\bea
\bar{\alpha}_0\bar{\alpha}_1\bar{\alpha}_2\bar{\alpha}_3=
\alpha_{0}^{*}\alpha_{1}^{*}\alpha_{2}^{*}\alpha_{3}^{*}\,-\,
\alpha_0\alpha_1\alpha_2\alpha_3\,.
\label{ayay}
\eea
There are some evident symmetries in the solution (\ref{fotwet}). For
instance, the free energy (\ref{fotwet}) is a symmetric function with
respect to arbitrary permutations of $\alpha_0, \alpha_1,
\alpha_2,\alpha_3$ parameters. We can as well change the signs of any two
of them, with $-\beta f_Q$ unchanged. There is also a less evident hidden
symmetry in the solution, corresponding to the Kramers-Wannier duality in
the case of the standard rectangular lattice.  Namely, the partition
function $Q\{\alpha_{}\}$ is invariant under the transformation $\alpha_{j}
\leftrightarrow \alpha_{j}^ {\,*}$. This symmetry in fact holds already for
the parameters of the separable fermionic modes (\ref{amodes}), and can be
proved making use of (\ref{aiden}), for further details see [19].

In order to establish the possible critical points, we have to look for
zeroes of $Q^{\,2}\,(p\,|\,q)$ momentum modes. As it can be guessed
already from the analogy between (\ref{amodes}) and (\ref{aiden}), there
are four exceptional modes with $(p\,|\,q)=\, (0\,|\,0), (0\,|\,\pi),
(\pi\,|\,0), (\pi\,|\, \pi)$. For these modes we have, respectively,
$Q^{\,2}\, (p\,|\,q)= (2\bar{\alpha}_0)^2\,,  \;(2\bar{\alpha}_1)^2\,,\;
(2\bar{\alpha}_2)^2\,, \;(2\bar{\alpha}_3)^2\,$. Remember that the
parameters $\alpha_{j}^{}$ and hence $\bar{\alpha}_j$ are some functions of
temperature. Thus, if at some temperature one of the above momentum modes
vanishes, we fall at the point of phase transition. It can be shown that
all other modes $Q^{\,2}(p\,|\,q)$ are definitely positive at all
temperatures, there are no other critical points for physical values of the
$\alpha$-parameters.  The possible criticality conditions can be combined
into one equation:
\bea
\bar{\alpha}_0\,\bar{\alpha}_1\,\bar{\alpha}_2\,\bar{\alpha}_3\;=\;0\,.
\label{acrit0}
\eea
It can be shown also that if all bond interactions are ferromagnetic
then $\bar{\alpha}_1, \bar{\alpha}_2, \bar{\alpha}_3$ never become zero,
and the critical point can only be associated with $\bar{\alpha}_0=0$.
Thus, for the purely ferromagnetic interactions the unique critical pont is
definite by the equation:
\bea
2\,\bar{\alpha}_0 = \alpha_0- \alpha_1 -\alpha_2 -\alpha_3 =0\,.
\label{acrit01}
\eea
The criticality conditions with $\bar{\alpha}_k =0$, $k=1,2,3$, can only be
realized when the antiferromagnetic interactions are involved in the
hamiltonian.

If the critical point is associated with $\bar{\alpha}_j=0$, then the
singular part of the free energy (\ref{fotwet}) near $T_c$ is given as
follows [19]:
\bea
-\beta f_Q\,|\,_{\rm singular }^{} = \frac{(2\bar{\alpha}_j)^2}
{16\pi\sqrt{(\alpha_0\alpha_1\alpha_2\alpha_3)_c}}
\,\ln\,\frac{\rm const}{(2\bar{\alpha}_j)^2}\;,
\label{fsing1}
\eea
which implies that, near $T_c$, in the order of magnitude $-\beta f_Q \sim
\tau^2\,\ln \tau^{\,2}$, where $\tau \sim |T-T_c|$. The specific heat
have thus the log-type singularity, $C \sim |\,\ln\tau\,|$ as $T
\rightarrow T_c$ (Onsager, 1944).

It is seen that the eigenvalues (\ref{aeigen}) play important role, but it
is not so clear, in fact, how the role of the polynomial $F_{123}$ can be
understood in a less formal way, at the level of the original spin-variable
formulation of the problem, prior to the analytic solution. The special
role which the parameters (\ref{aeigen}) play in the inherent structure of
the 2D Ising model is confirmed by the expression for the spontaneous
magnetization. The 8-th power of the spontaneous magnetization
$M=\left<\sigma_{mn}\right>$ for model (\ref{q123}) is given by the
following very simple expression [19]:
\bea
M^{\,8}\,=\, (-1)\,
\frac{\bar{\alpha}_0\bar{\alpha}_1\bar{\alpha}_2\bar{\alpha}_3}
{\alpha_0\alpha_1\alpha_2\alpha_3}\,=\,1\,-\,
\frac{\alpha_{0}^{*}\alpha_{1}^{*}\alpha_{2}^{*}\alpha_{3}^{*}}
{\alpha_0\alpha_1\alpha_2\alpha_3}\,.
\label{magn}
\eea
This expression for $M^{\,8}$ holds true when the right hand side varies
between 0 and 1, and $M^8=0$ otherwise. It is easy to check that the known
expressions for the spontaneous magnetizations of the rectangular,
triangular, and hexagonal lattices follow easily from (\ref{magn}) as
particular cases. From (\ref{magn}) we find $M \sim\tau^{ \frac{1}{8}}$ as
$\tau \sim |\,T-T_c\,|\,\to0$, with the universal value of the critical
index $\beta=1/8$ for the magnetization at the critical isobar. What are
the hidden reasons for such simple expression for $M^{\,8}$, this is yet
unknown.

%
%
\section{ Majorana fields.}

The fermionic integral for $Q$ given in (\ref{qccc}) also appears to be a
suitable starting point to formulate the continuum limit field theories for
the correspondent models near $T_c$ \cite{icsmp95}. For other approaches to
the contnuum-limit formulation of the 2D Ising model also see
\cite{zubiz77}, \cite{droiz89}, \cite{shal94}. Assuming the purely
ferromagnetic case, with $\alpha_0, \alpha_1, \alpha_2, \alpha_3$ all
positive, we write once again the exact lattice action from (\ref{qccc}) as
follows:
\bea
S\,(c)\,= \sum\limits_{mn}\;\big[\,(\alpha_0-\alpha_1-\alpha_2
-\alpha_3)\,c_{mn}^{}c_{mn}^{*} +\,\alpha_1\,c_{mn}^{}\,(c_{mn}^{*}-
c_{m-1n}^{*})\,
\cr
+\,\alpha_2\,c_{mn}\,(c_{mn}^{*}- c_{mn-1}^{*}) +\,\alpha_3\,
c_{mn}\,(c_{mn}^{*}-c_{m-1n-1}^{*})\,
\cr
+\,\alpha_1\,c_{mn}^{}\,(c_{mn}^{}-c_{m-1n}^{}) +\,\alpha_2\,c_{mn}^{*}
(c_{mn}^{*}-c_{mn-1}^{*} )\,\big]\,,
\label{sccc0}
\eea
with $\,c_{mn}^{\,2}=c_{mn}^{\,*\;2}=0\,$. Let us define the lattice
derivatives as follows: $\partial_m\,x_{mn}=x_{mn}-x_{m-1n}\,,\, \partial_n
\,x_{mn}=x_{mn}- x_{mn-1}\,$, also note that  $x_{mn}-x_{m-1n-1}=
\partial_m\,x_{mn} + \partial_n\,x_{mn}-\partial_m \partial_n\,x_{mn}\,$.
Introducing the new notation for the fermionic fields, $c_{mn}\,,\,
c_{mn}^{*} \to \psi_{mn}\,,\, \bar{\psi}_{mn}\,$, we find the action
(\ref{sccc0}) in the form:
\bea
S\,(\psi) =\sum\limits_{mn}^{}\,\big[\;
\underline{m}\,\psi_{mn}\bar{\psi}_{mn} +\,\lambda_1\,\psi_{mn}\,
\partial_m\, \bar{\psi}_{mn}+
\,\lambda_2\,\psi_{mn}\,\partial_n\, \bar{\psi}_{mn}
\cr
\,-\,\alpha_3\,\psi_{mn}\,\partial_m\,\partial_n\,\bar{\psi}_{mn} +
\alpha_1\,\psi_{mn}\,\partial_m\,\psi_{mn} +
\alpha_2\,\bar{\psi}_{mn}\, \partial_n\,\bar{\psi}_{mn}\,]\,,
\label{spsi0}
\eea
\bea
\lambda_1=\alpha_1+\alpha_3,\;\; \lambda_2=\alpha_2+\alpha_3\,,\;\;\;
\underline{m}=\alpha_0 -\alpha_1 -\alpha_2 -\alpha_3= 2\,\bar{\alpha}_0\,.
\nonumber
\eea
It is easy to recognize in this still exact lattice action the typical
relativistic field-theoretical like structure with mass term and kinetic
part.  Since we assume the ferromagnetic case, the criticality condition
is $\,2\,\bar{ \alpha}_0=0\,$, see (\ref{acrit01}). The parameter
$\underline{m}= \alpha_0- \alpha_1- \alpha_2-\alpha_3 =
\,2\,\bar{\alpha}_0\,$ plays the role of mass in the field-theoretical
interpretation. The mass vanishes at the critical point, $\,\underline{m}
\sim|\,T-T_c\,|\to 0\,$ as $T\to T_c\,$. The ordered phase corresponds to
$\underline{m}<0\,$.

Taking in (\ref{spsi0}) the formal limit to the continuum euclidean space,
with $(mn)\to x=(x_1\,|\,x_2)$, $\,\psi_{mn} \to \psi(x)= \psi\,(x_1\,|\,
x_2)$, and $\partial_m \to \partial_1=\partial/\partial x_1$, $\partial_n
\to \partial_2= \partial/\partial x_2\,$, and neglecting the second-order
kinetic term with $\partial_1\partial_2$, we obtain the Majorana like
fermionic action of the correspondent continuum theory:
\bea
S = \int d^2x \;[\,\underline{m}\,\psi(x)\ovl{\psi}(x) +
\lambda_1\, \psi(x)\,\partial_1\,\ovl{\psi}(x) +
\lambda_2\, \psi(x)\,\partial_2\,\ovl{\psi}(x)
\cr
\,+\,\alpha_1\, \psi(x)\,\partial_1\,\psi(x) +
\alpha_2\, \ovl{\psi}(x)\,\partial_2\,\ovl{\psi}(x)\,]\,,
\label{maj1}
\eea
where the parameters $\lambda_1$, $\lambda_2$, $\underline{m}$ are defined
in (\ref{spsi0}). A nonstandard feature in this action are the
"nondiagonal" kinetic terms $\psi\partial_1\bar{\psi}$, $\psi\partial_2
\bar{\psi}$.  These terms can be eliminated by a suitable linear
transformation of the fields: $\psi= \gamma\psi_1 + \eta\psi_2,\,
\bar{\psi}= \bar{\gamma}\psi_1+ \bar{\eta}\psi_2$, where $\psi_1, \psi_2$
are new Majorana components. The condition that the nondiagonal like terms
will not appear in (\ref{maj2}) is $\gamma\bar{\gamma}= \eta\bar{\eta}$.
Omitting further details, after a suitable transformation of the fields and
the axis in the $d^{\,2}x$ space, from (\ref{maj1})
we obtain Majorana action in the canonical form
(cf.~\cite{zubiz77,droiz89}):
\bea
S = \int d^2x \;\big[\,\overline{m}\,\psi_1(x)\,\psi_2(x) +
\psi_1(x)\,\pal_0\,\psi_1(x) -\psi_2(x)\,\ovl{\pal}_0\,\psi_2(x)\,\big]\,,
\label{maj2}
\\  \nonumber
\pal_0= \frac{1}{2}\,(\pal_1+i\pal_2)\,,\;\;\ovl{\pal}_0=
\frac{1}{2}\,(\pal_1-i\pal_2)\,, \;\;
\mbox{2D Majorana}\,,
\eea
with the rescaled mass:
\bea
\ovl{m}= \frac{\alpha_0 -\alpha_1-\alpha_2 -\alpha_3}
{\big(2\,\sqrt{(\alpha_0\alpha_1\alpha_2\alpha_3)_{c}}\big)^{\,1/2}}\;.
\label{mass2}
\eea
The 2D Ising model is presented now as a field theory of free massive
two-component Majorana fermions in euclidean $d^{\,2}x$ space.  The new
Majorana fields in this representation, $\,\psi_1\,, \,\psi_2\,$, are the
linearly transformed fields $\,\psi\,,\,\bar{\psi}\,$ from (\ref{maj1}).
The axis of the $d^{\,2}x$ space are also rescaled and rotated as we pass
>from (\ref{maj1}) to (\ref{maj2}). Respectively, the mass is rescaled
according (\ref{mass2}), here $(\,)_c$ means the criticality condition
$(\alpha_0- \alpha_1 -\alpha_2-\alpha_3)_c =0$.

In matrix notation, the Majorana action (\ref{maj2}) becomes:
\bea
S_{\rm\,major} = \frac{1}{2}\,\int d^{\,2}x\,
\left(\ba{c} \psi_{1}^{} \cr \psi_{2}^{} \ea\right)
\left[\, \ovl{m}\left(\ba{rl} 0 & 1 \cr -1 & 0 \ea\right) +
\left(\ba{cc}
\pal_1+ i\,\pal_2&0 \cr 0&-\pal_1 + i\,\pal_2 \ea\right)
\right]\left(\ba{c} \psi_{1} \cr \psi_{2} \ea\right)\,.
\label{maj3}
\eea
Introducing the standard Pauli matrices:
\bea
\sigma_1=\left(\ba{lr} 0&1 \cr 1&0 \ea\right)\,,  \triex
\sigma_2=\left(\ba{lr} 0&-i \cr i&0 \ea\right)\,, \triex
\sigma_3=\left(\ba{lr} 1&0 \cr 0&-1 \ea\right)\,,   \triex
\label{pauli1}
\eea
the action (\ref{maj3}) can be written also in the form:
\bea
S_{\rm\,major} = \frac{1}{2}\,\int d^{\,2}x\;\Psi^{}\,(x)\,
[\,\overline{m}\,(i\,\sigma_2)\, + \pal_1\,(\sigma_3) +
\,i\,\pal_2\,(1)\,]\;\Psi^{}\,(x)\;, \triex
\Psi\,(x)= \left(\ba{c} \psi_1 \cr \psi_2 \ea\right),
\label{maj4}
\eea
or in the form:
\bea
S_{\rm\,major} = \frac{1}{2}\,\int d^{\,2}x\;\tilde{\Psi}\,(x)\,
[\,\overline{m}\,
+ \pal_1\,(\sigma_1) + \pal_2\,(\sigma_2)\,]\;\Psi^{}\,(x)\;, \triex
\tilde{\Psi}\,(x) = \Psi\,(x)\,(i\,\sigma_2)\,.
\label{maj5}
\eea
Introducing the 2D gamma-matrices in a natural way: $\gamma_1=\sigma_1$,
$\gamma_2= \sigma_2$, the action (\ref{maj5}) becomes:
\bea
S_{\rm\,major} = \frac{1}{2}\,\int
d^{\,2}x\;\tilde{\Psi}\,(x)\,[\;\overline{m}\, +
\hat{\pal}\,]\;\Psi^{}\,(x)\;, \triex
\hat{\pal}= \gamma_1\,\pal_1 + \gamma_2\,\pal_2\,,
\triex  \hat{\pal}^{\,2} = \pal_{1}^{\,2} +\pal_{2}^{\,2}\,,
\label{maj6}
\eea
which is the 2D Majorana action written in the standard form assumed in
relativistic field theory.  Notice that the conjugated Majorana spinors
$\tilde{\Psi}$ and $\Psi$ here are not the truly independent fields since
they both are built from the same component fields $\psi_1$, $\psi_2$. The
truly independent left and right spinors $\bar{\Psi}$ and $\Psi$ appear if
pass to the Dirac theory. This is commented in next section.

%
%
\section{ Dirac fields.}

We can pass to the Dirac field theory of charged fermions by doubling the
number of fermions in the Majorana representation
(\ref{maj2})-(\ref{maj6}). On formal level, this corresponds to passing
>from Pfaffian like Gaussian integral to the determinantal Gaussian integral
according the identity $|\Pfaff\,A\,| ^{\,2}=\det A$, see (\ref{detpf}). To
this end, we take the two identical copies $S'$ and $S''$ of
the Majorana action (\ref{maj6}) and consider the combined action
$S_{\rm\,dirac}=S'+S''$. In this combined action we introduce the new
Dirac fermionic fields $\Psi = (\psi_{1}\,|\,\psi_{2})$ and $\ovl{\Psi}=
(\psi_{1}^{\,*}\,|\,\psi_{2}^{\,*})$ by substitution:
\bea
\Psi = \frac{1}{\sqrt{2}}\,
\big(\Psi^{\,'} + i\,\Psi^{\,''}\big)\,, \triex
\ovl{\Psi} = \frac{1}{\sqrt{2}}\,
\big(\tilde{\Psi}^{\,'} - i\, \tilde{\Psi}^{\,''}\big)\,,
\label{subst1}
\eea
where $\Psi{\,'}, \Psi^{\,''}, \tilde{\Psi}^{\,'}, \tilde{\Psi}^{\,''}$
are the original Majorana spinors.  In terms of the new fields the combined
action $S_{\rm\,dirac} = \,(S^{\,'} + S^{\,''})_{\rm\,majorana}$ becomes:
\bea
S_{\rm\,dirac} = \,\int d^{\,2}x\;\ovl{\Psi}\,(x)\,[\,\overline{m}\, +
\hat{\pal}\,]\;\Psi^{}\,(x)\;, \triex
\hat{\pal}= \gamma_1\,\pal_1 + \gamma_2\,\pal_2\,,
\label{dirac1}
\eea
where the gamma matrices are the same as in (\ref{maj6}). If written in
components, the substitution inverse to (\ref{subst1}) takes the form:
\bea
\psi_{1}^{\,'}\,=  \frac{1}{\sqrt{2}}(\psi_1+ \psi_{2}^{\,*})\,, \triex
\,i\,\psi_{1}^{\,''}\,= \frac{1}{\sqrt{2}}(\psi_1- \psi_{2}^{\,*})\,,
\nonumber \\
\psi_{2}^{\,'} = \frac{1}{\sqrt{2}}(\psi_2- \psi_{1}^{\,*})\,, \triex
\,i\,\psi_{2}^{\,''} = \frac{1}{\sqrt{2}}(\psi_2+ \psi_{1}^{\,*})\,.
\label{subst2}
\eea
The four components of the Dirac spinors $\ovl{\Psi}$, $\Psi$ are all
independent variables. By convention, we can assume $\psi_{1}^{\,*},
\psi_{2}^{\,*}$ to be complex conjugated to $\psi_1, \psi_2$. In this case
$\ovl{\Psi}$ is the hermitian conjugated to $\Psi$. The advantage of the
transformation from Majorana to Dirac formulation is that we gain in later
case the gauge symmetry which can be useful in some cases \cite{shal94}.
The continuum-limit formulations like (\ref{maj6}) and (\ref{dirac1})
captures all relevant features of the exact lattice theory at low
momenta, or at large space scales, which is only important for the
the critical-point singularities in thermodynamic functions near $T_c$.
For the simplest case of the standard rectangular lattice, the continuum
limit was used to study the behaviour of the 2DIM spin  correlation
functions at large separation in \cite{zubiz77,droiz89}.  Another
important field of application of the continuum-limit formulation are the
Ising models with quenched disorder [20-23].

In conclusion we note that the singular part of the free energy and
specific heat can be extracted, in the exact form, from (\ref{maj6})
and/or (\ref{dirac1}). For instance, starting from (\ref{dirac1}), we
find:

\bea
-\beta\,f_{\rm\,sing}^{} = \frac{1}{2}\,\ln\,Q^{\,2} =
\frac{1}{2}\,\int d^2x\,\ln\,Q^{\,2}(x) =
\frac{1}{2}\,\int\frac{d^2p}{(2\pi)^2}\,\ln\,Q^{\,2}(p)
\cr
= \frac{1}{8\pi^2}\,\int d^2p\,\ln(\overline{m}^2 + p^2) =
\frac{\pi}{8\pi^2}\,\ovl{m}^2\ln\,\frac{\rm\,const}{\ovl{m}^2}
\,+\,(...)\;.
\label{fsing1dir}
\eea
This is equivalent to the equation (\ref{fsing1}) obtained
directly from the exact expression for the lattice free energy
(\ref{fotwet}), taking into account the definition of the effective mass
given in (\ref{mass2}).

%
%
\section{ Conclusions.}

We have discussed some aspects of a simple fermionic interpretation of the
2D Ising models in terms of the anticommuting integrals. For any planar 2D
Ising model, the partition function can be expressed as a fermionic
Gaussian integral. The analytic solutions for regular lattices then easily
follow by transformation to the momentum space. The continuum limit
field-theoretical formulations for the 2D Ising models also can be deduced
readily from the exact lattice fermionic integral for the partition
function. The differences between particular lattices are merely adsorbed,
in the field-theoretical limit, in the definition of the effective mass.
Finally, the free-fermion interpretation for the 2D Ising model shows that
this model, which in its original formulation is rather discrete
combinatorial problem, can be placed, in fact, into a common range
of the typical models of quantum statistics and solid state physics.

%
%
%

\end{document}